\documentclass[manuscript]{aastex}

\shorttitle{Mg-depleted extrasolar terrestrial planets}
\shortauthors{Carter-Bond et al.}

\begin{document}

\title{Low Mg/Si planetary host stars and their Mg-depleted terrestrial planets}

\author{Jade C. Carter-Bond\altaffilmark{1,2} and David P. O'Brien}
\affil{Planetary Science Institute, 1700 E. Fort Lowell, Tucson, AZ 85719}
\email{j.bond@unsw.edu.au}

\and

\author{Elisa Delgado Mena and Garik Israelian}
\affil{Instituto de Astrof\'{\i}sica de Canarias, 38200 La Laguna, Tenerife, Spain}
\affil{Departamento de Astrof\'{\i}sica, Universidad de La Laguna, 38205 La Laguna, Tenerife, Spain}

\and

\author{Nuno C. Santos}
\affil{Centro de Astrof\'{\i}sica, Universidade do Porto, Rua das Estrelas, 4150-762 Porto, Portugal}
\affil{Departamento de F\'{\i}sica e Astronomia, Faculdade de Ci\^{e}ncias, Universidade do Porto, Portugal}

\and

\author{Jonay I. Gonz\'alez Hern\'andez}
\affil{Instituto de Astrof\'{\i}sica de Canarias, 38200 La Laguna, Tenerife, Spain}
\affil{Departamento de Astrof\'{\i}sica, Universidad de La Laguna, 38205 La Laguna, Tenerife, Spain}

\altaffiltext{1}{Now at Department of Astrophysics, School of Physics, University of NSW, NSW 2052, Australia}
\altaffiltext{2}{Formerly published as Jade C. Bond}

\begin{abstract}
Simulations have shown that a diverse range of extrasolar terrestrial planet bulk compositions are likely to exist, based on the observed variations in host star elemental abundances. Based on recent studies, it is expected that a significant proportion of host stars may have Mg/Si ratios below 1. Here we examine this previously neglected group of systems. Planets simulated as forming within these systems are found to be Mg-depleted (compared to the Earth), consisting of silicate species such as pyroxene and various feldspars. Planetary carbon abundances also vary in accordance with the host stars C/O ratio. The predicted abundances are in keeping with observations of polluted white dwarfs, lending validity to this approach. Further studies are required to determine the full planetary impacts of the bulk compositions predicted here.
\end{abstract}


\section{Introduction}

Extrasolar planetary host stars are well known to display systematic enrichments in many elements, including several key terrestrial planet forming elements such as Fe, Mg, Si and O (e.g. \citealt{g3,g6,sb,gilli,ec,oxygen}). Furthermore, a wide range in host star photospheric Mg/Si and C/O values\footnote{Note this is the elemental ratio, not the Solar normalized logarithmic ratio, which is commonly shown as [X/H].} has been observed \citep{petigura:2011,brugamyer:2011,delgado-mena:2010}. The ratios of Mg/Si and C/O are critical for planetary systems as they determine the bulk mineralogy of the solid, planet forming material present within the disk, thus also controlling the composition of any terrestrial planets present within the system. Assuming equilibrium conditions, a C/O value greater than 0.8 implies that Si will exist in the solid form as SiC, with additional C also present. For C/O values less than 0.8, silicates are produced with either the SiO$_{4}$$^{4-}$ or SiO$_{2}$ building block. The exact composition of these silicates is controlled by the Mg/Si value, ranging from pyroxene (MgSiO$_{3}$) and various feldspars (for Mg/Si$<$1), to a combination of pyroxene and olivine (Mg$_{2}$SiO$_{4}$) (for 1$<$Mg/Si$<$2) and finally to olivine with other MgO or MgS species (for Mg/Si$>$2). Note that for reference, the solar Mg/Si value is 1.05 \citep{asp}, while the bulk Earth Mg/Si value is 1.02 \citep{kandl}, resulting in planets composed of both olivine and pyroxene. The observed variations in these key ratios for known planetary host stars implies that a wide variety of extrasolar terrestrial planet compositions are likely to exist, ranging from relatively ``Earth-like" planets to those that are dominated by C as graphite and carbide phases (e.g. SiC, TiC) \citep{bond:2010b,kuchner:2005}.

Previous studies have examined several possible terrestrial planet compositions based upon stellar photospheric abundances \citep{bond:2010b}. Recent work by \cite{delgado-mena:2010} has shown that a significant fraction of host stars (56\%) may have a Mg/Si value less than 1. Such compositions have not been previously simulated and are expected to differ significantly from that of Earth. Given the large errors associated with photospheric Mg/Si (and C/O) values, the true fraction of host stars with Mg/Si values below 1 is likely to be less than 56\%. However, it is not expected to be negligible, implying that a significant number of terrestrial extrasolar planets may have compositions unlike that of Earth.

In this Letter we present the results of simulations of terrestrial planet formation within three extrasolar planetary systems with stellar abundances determined by \cite{delgado-mena:2010} to have Mg/Si values less than 1. The focus of this letter is a new region of Mg/Si and C/O space that was not previously occupied by any known planetary host stars, hence the very specific limitations of the simulations conducted here. The systems studied highlight the main chemical differences between these systems and those previously studied. We adopt the same approach as used in \cite{bond:2010a,bond:2010b}, combining dynamical simulations of terrestrial planet formation with chemical equilibrium models of the composition of solid material within the disk. This is the first study to consider terrestrial planets of this nature, assisting in developing our understanding of the full spectrum of possible extrasolar terrestrial planet commotions.

\section{Simulations}
\subsection{Extrasolar Planetary Systems}
Three known extrasolar planetary systems with Mg/Si values less than 1 were selected for this study. Each of the three systems was previously simulated by \cite{bond:2010b} with different stellar abundances. The stellar elemental abundance values of \cite{delgado-mena:2010} were utilized in this study as they are the first to determine the abundance of all of the required elements in a completely internally consistent manner (high quality spectra and identical approach for all stars and elements) for a large sample of both host and non-host stars. It should be noted that although one system selected for study here (55Cnc) was not observed as part of the HARPS sample, it was observed as part of the CORALIE survey. Spectral analysis was completed in the same fashion as for all other HARPS target stars, thus maintaining the consistency of the methodology applied here. Instrument variation is not thought to have introduced significant error. All three systems were previously found to have Mg/Si values greater than 1 and C/O values above 0.8 (based on \cite{gilli,be,ec,oxygen}) (see Table \ref{input_chem} for both previous and current Mg/Si and C/O values). Abundance variations are understood to be due to differences in atomic and stellar parameters used, expanded line lists and improved equivalent width measurements. Of the three systems to be studied here, two host stars were found by \cite{delgado-mena:2010} to have C/O values less than 0.8 (HD17051 and HD19994), while one host star (55Cnc) has a C/O value well above 0.8. This range in C/O values was purposefully chosen to examine the full range of planetary compositions possible with Mg/Si values below 1. As with all such abundance studies, the error bars associated with these values are large and cannot be minimized with current techniques. With this sample, we are also examining a variety of planetary system architectures. Both HD17051 (HR810, HIP12653, iota Hor) and HD19994 have a single known giant planetary companion with masses slightly larger than that of Jupiter (2.26M$_{Jupiter}$, a=0.925AU, e=0.161 and 1.7M$_{Jupiter}$, a=1.42AU, e=0.3, respectively). On the other hand, 55Cnc (HD75732) has five known planetary companions (4 inner planets with a$\leq$0.781AU and one outer planet with a=5.76AU) with masses ranging from 3.84M$_{Jupiter}$ down to 0.03M$_{Jupiter}$ (8.58M$_{\bigoplus}$).

\subsection{Chemical Simulations}
In this study, we utilized the same approach as in \cite{bond:2010a,bond:2010b}. The composition of solid material within the disk is assumed to be in equilibrium with the primordial stellar nebula. Consequently, equilibrium condensation sequences can be used as a proxy to determine the final elemental composition of the simulated terrestrial planets. Following \cite{bond:2010a,bond:2010b}, condensation sequences for the 16 major solid forming elements (H, C, N, O, Na, Mg, Al, Si, P, S, Ca, Ti, Cr, Fe and Ni) were obtained using HSC Chemistry (v. 5.1) and are based upon the Gibbs energy minimization method. Stellar photospheric abundances of Fe, Mg, Si, C, O, Ni (from \cite{delgado-mena:2010}), Na, Al, Ca, Ti and Cr (derived in this work using the linelist from \cite{neves:2009}) were utilized as the composition of the stellar nebula. These abundances were determined in a uniform, internally consistent manner, thus reducing any possible systematic errors within the sample. Abundances of N, P and S were approximated based on the odd-even effect. The increased stability of even atomic number nuclei results in a higher occurrence of these species when compared to odd atomic number nuclei, producing the well known saw tooth pattern in solar abundances. There is no reason to think that this same trend would not hold for extrasolar planetary host stars. Thus by fitting a linear trend through the odd and even atomic number nuclei separately for solar abundances we were able to approximate the abundance of N, P and S based on the observed abundances of C, Al and Si (respectively). The input values used in HSC Chemistry (normalized to 10$^{6}$ Si atoms) for each system are shown in Table \ref{input_chem}. In order to provide a spatial location within the disk for a given composition, radial pressure and temperature profiles from \cite{hersant} were applied. Disk conditions at a evolutionary time of t = 5$\times$10$^{5}$yr are utilized here as they were found to produce the best fit to known planetary values within the Solar System \citep{bond:2010a}. For more detail on this method, the reader is referred to \cite{bond:2010a} and \cite{bond:2010b}.

\subsection{Dynamical Simulations\label{dyn}}
Numerous simulations of terrestrial planet formation have been conducted, focussing on aspects such as disk mass and viscosity \citep{thommes:2008}, giant planet migration \citep{mandell:2007} and orbital parameters \citep{raymond:2009}. For this study, we utilized the n-body simulations of late-stage terrestrial planet accretion of \cite{bond:2010b}. These simulations are intended to be indicative of the types of terrestrial planets that may form within the systems selected, not be inclusive of all aspects of planet formation. Four simulations were completed for each system. Each simulation was run using the SyMBA n-body integrator \citep{duncan} with orbital parameters of the giant planets taken from the catalog of \cite{but-cat} (updated from exoplanets.org). An initial population of Lunar-to Mars-mass embryos was distributed between 0.3 AU from the host star and the known giant planet (in the case of 55Cnc, embryos were distributed between the inner and outer planets) in accordance with the embryo mass, spacing and orbital radius relations of \cite{kandi} and following the MMSN solid surface density. As we are currently only considering late-stage accretion, migration is neglected in the current simulations. For more details, please refer to \cite{bond:2010b}. Giant planet migration is expected to alter the composition of the terrestrial planet feeding zones, thus also changing the composition of the final terrestrial planet. Simulations incorporating this effect are currently ongoing.

\subsection{Combining Dynamics and Chemistry}
As in previous work, the chemical and dynamical simulations were combined together by assuming each embryo retains the composition of its formation location and contributes the same composition to the simulated terrestrial planet. In this approach, phase changes and outgassing are neglected. Following the method of the dynamical simulations, all collisions are assumed to result in a perfect merger (i.e. no mass was lost from either the target body or the impactor during impact).

\section{Results and Discussion}
\subsection{Planetary Compositions}
\begin{figure}
\begin{center}
\includegraphics[width=120mm]{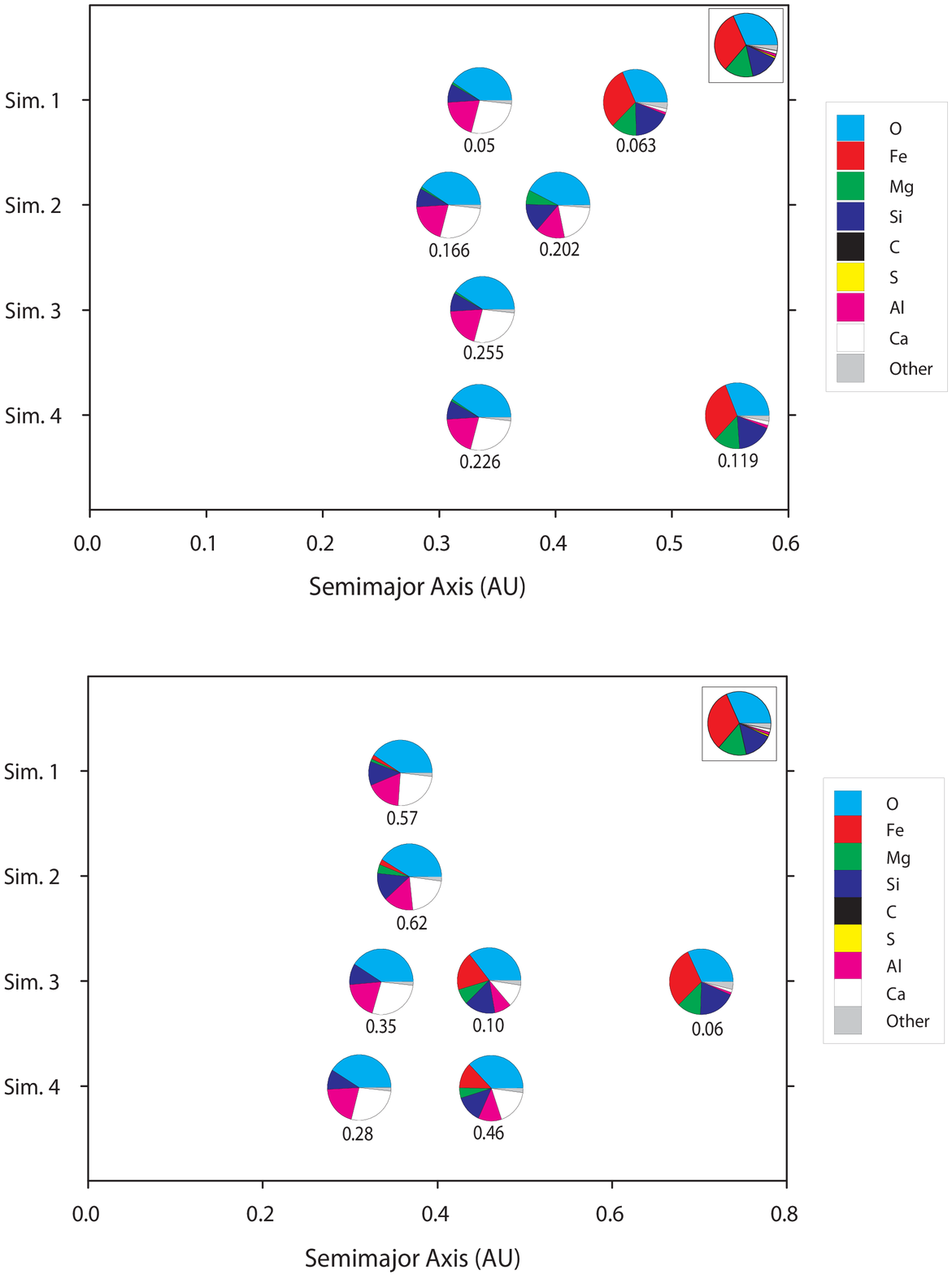}\caption[]
{Schematic of the bulk elemental planetary composition for systems with C/O$<$0.8: HD17051 (top) and HD19994
(bottom). All values are wt\% of the final simulated planet. Values are shown for the terrestrial planets produced in each of the four
simulations run for the system. Numbers refer to mass of planet in Earth masses. Earth values taken from \cite{kandl} are shown in the upper right of each panel. \label{allpie1}}
\end{center}
\end{figure}

\begin{figure}
\begin{center}
\includegraphics[width=120mm]{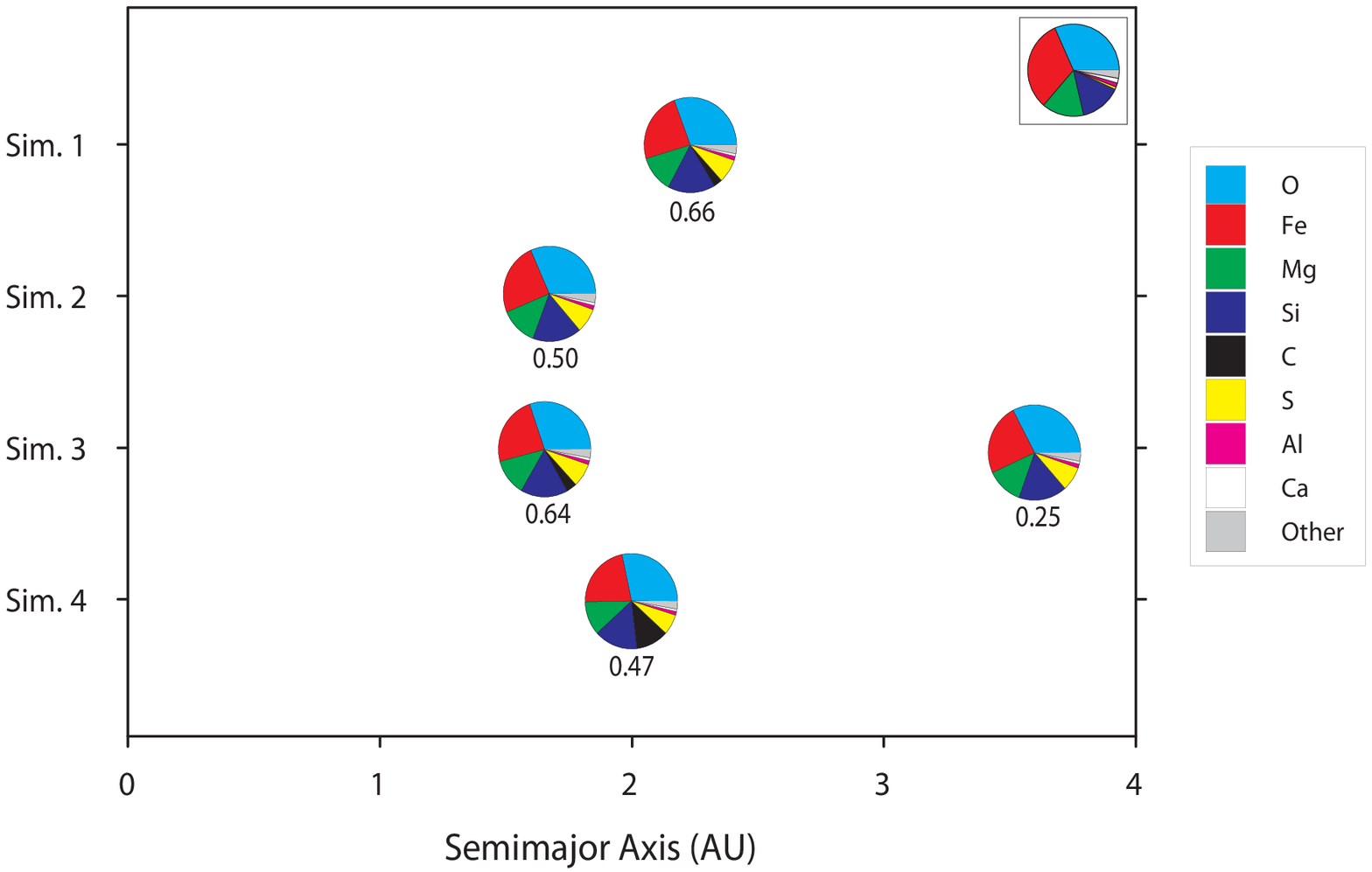}\caption[]
{Schematic of the bulk elemental planetary composition for 55Cnc (C/O$>$0.8). All values are wt\% of the final simulated planet. Values are shown for the terrestrial planets produced in each of the four
simulations run for the system. Numbers refer to mass of planet in Earth masses. Earth values taken from \cite{kandl} are shown in the upper right. \label{allpie2}}
\end{center}
\end{figure}

A schematic of the resulting bulk elemental compositions of the simulated terrestrial planets are shown in Figures \ref{allpie1} and \ref{allpie2}. Note that these compositions were produced using the radial disk profiles at t = 5$\times$10$^{5}$yr as previously discussed. These same results are shown numerically in Table \ref{results}. All of the terrestrial planets considered here have compositions dominated by O, Fe, Mg and Si with most of these elements being delivered in the form of silicates or metals (in the case of iron). However, important differences between those planets forming in systems with C/O$<$0.8 (HD17051, HD19994) and those with C/O$>$0.8 (55Cnc) can be seen.

Although all of the simulated planets are composed of Mg-silicates and Fe, only 2 of 7 simulated planets for HD17051 and 1 of 7 simulated planets for HD19994 contains sufficient Mg to be classed as ``Earth-like". Note that here an ``Earth-like" planet does not mean that a planet is identical in compositional identical Earth. Rather, it has a broadly similar composition, consisting of Mg silicates and metallic Fe with other species present in relatively minor amounts and with a maximum deviation of $\pm$25\% from the elemental abundances for the Earth listed in \cite{kandl} for the major elements (O, Fe, Mg, Si). This variation limit was selected so that both Venus and Mars would be considered to be ``Earth-like" in composition.

All other planets produced in HD17051 and HD19994 are considered to be Mg-depleted silicate planets. Furthermore, as in prior simulations, radial variations can be seen for those simulations producing multiple terrestrial planets (3 simulations for HD17051, 2 simulations for HD19994). In these cases, the innermost terrestrial planets (located within $\sim$0.5 AU from the host star) contain a significant amount of the refractory elements Al and Ca ($\sim$47\% of the planetary mass). Planets forming beyond $\sim$0.5 AU from the host star contain steadily less Al and Ca with increasing distance. With the inclusion of giant planet migration in future simulations, it is expected that this radial compositional gradient will diminish, leaving us with essentially an averaged composition more closely resembling that of Earth. However, for the current simulations and with the exception of the three ``Earth-like" planets previously mentioned, these systems can be described as producing refractory-enriched silicate planets for the interior planets and Mg-depleted silicate planets for the outermost planets. None of the planets accrete any water or other hydrous phases in these simulations due to their distance from the snow line and narrow feeding zones.

Only 1 system studied here has a C/O ratio above 1 (55Cnc, C/O = 1.12). This system produced carbon-enriched ``Earth-like" planets, based on the above definition. Although the system is predicted to contain a region between approximately 0.46 and 1.48AU in which the solid composition is almost exclusively C (present as graphite, SiC and TiC), all of the simulated planets formed outside of this region (between 1.6 and 3.6AU) due to the inner planets of this system. Thus, instead of being dominated by C phases, the planetary feeding zones contained both C and significant amounts of other Mg silicates (such as pyroxene). As such, the simulated planets are Earth-like in their bulk compositions, with C contents ranging from $<$0.01 wt\% to 11 wt\%. This implies that the innermost giant planets should contain significant amounts of C as they lie well within this C-dominant zone (assuming that they also formed in or close to their current orbits or sourced a significant amount of material from this region). Additionally, given that the location of this C-rich zone varies with time as the disk cools, simulations using a different radial pressure and temperature profile may produce C-rich terrestrial planets. Likewise, the planets simulated within the 55Cnc system are enriched in S compared to the HD17051 and HD19994. This is not due to S enrichment within the system itself but rather due to the feeding zones for the simulated planets being further from the host star and thus cool enough to contain S in the solid form (primarily as FeS). Giant planet migration is expected to drastically alter this marked radial variation in composition and is the focus of current simulations. As in the other two systems, all of the simulated planets within 55Cnc form `dry' in that they do not accrete any hydrous phases.

\subsection{Accretion Observations}
Although detections have been made of planets with masses as low as 2M$_{\bigoplus}$, we are still yet to detect planets with masses $\leq$1M$_{\bigoplus}$, such as those produced by the current simulations. As such, direct comparison between simulations and observations is not yet possible. However, we do have another avenue currently available to us through the observation of polluted white dwarfs. Pollution of the white dwarf photosphere is thought to be produced by the accretion of solid material from a planetary system orbiting the progenitor star. As the host star evolves and undergoes mass loss, the planetary system is disrupted, producing dynamically unstable orbits and resulting in the subsequent accretion of planetary material onto the white dwarf \citep{jura:2008,jura:2003}. As a white dwarf atmosphere is composed of only H and/or He, any accreted material can be detected, provided it is observed before it disperses throughout the convective zone.

Recent studies \citep{zuckerman:2011,klein:2011,klein:2010,zuckerman:2010,dufour:2010} have detected evidence of the accretion of rocky planetesimals onto white dwarfs. Composed of Mg, Si, O and Fe, the accreted material is similar to Earth in bulk composition. \cite{zuckerman:2011} observed Mg/Si values in the white dwarf NLTT 43806 of 1.05-1.26, consistent with accretion of Earth-like material. Furthermore, \cite{klein:2011} observed pollution that may be consistent with the accretion of a refractory-rich body, such as those produced in the inner regions of HD17051 and HD19994 by the present simulations. These observations appear to support the results of this study in that extrasolar terrestrial planets with Mg-silicate compositions are likely to exist, with refractory-rich ``Earth-like" bodies also possible (although not confirmed). Polluted white dwarf surveys including carbon have not yet been completed, preventing us from comparing our C-rich simulated planets to direct observations.

\subsection{Planetary Interiors and Processes}
Several broad inferences about the nature of the simulated extrasolar terrestrial planetary interiors can be made on the basis of the current simulations by assuming crystal settling in a global magma ocean is driven by density (i.e. most dense species sink to the core, least dense species float as a crust). The interiors for the Mg-depleted silicate planets are expected to contain an Fe-Ni-S core, overlain by a mantle of diopside (CaMgSi$_{2}$O$_{6}$) and spinel (MgAl$_{2}$O$_{4}$), with a crust of pyroxene and feldspar (NaAlSi$_{3}$O$_{8}$). On the other hand, due to their relatively low carbon enrichment, the planets of 55Cnc will have a Fe-Ni-S core with a spinel, diopside and olivine mantle and a pyroxene, feldspar and graphite crust. Given the low mass of all of the simulated planets (M$\leq$0.67M$_{\bigoplus}$), unless a significant amount of radioactive material is accreted by the planet or tidal heating is significant, it will be difficult to produce substantial amounts of magmatic melt. Any melt that is produced is expected to be intermediate to felsic in composition due to the lower abundance of Mg. However, given the high content of refractory elements, production of significant amounts of melt within the mantle would be severely limited.

Simulations of the interior processes of super Earth planets have been completed (e.g. \citealt{valencia:2007,oneill:2007}) assuming a planetary composition similar to that of Earth. In light of the range of planetary compositions now predicted to exist and the implications that the compositional variations may have on a planetary interior, such simulations should consider a more diverse range of compositions.

\subsection{Terrestrial Planet Searches}
As all of the systems examined in this Letter are capable of producing Mg-silicate planets, albeit with non-Earth-like compositions, systems orbiting host stars with a Mg/Si ratio less than 1 should still be included in terrestrial planet searches. Due to their predominantly silicate composition, terrestrial planets within these systems are not expected to be overly dark. Their albedo should be comparable to light C and S type asteroids (albedo $\sim$ 0.10-0.20), depending on the distribution of carbon (if any) within the crust. Thus it is theoretically possible that planets such as those simulated here may be detected via reflected light.

\section{Summary}
Recent observational studies have shown that a significant proportion of planetary host stars may have an Mg/Si ratio less than 1. We have simulated terrestrial planet formation within three such systems and found that such systems produce a variety of planetary compositions, depending on the location of the planetary feeding zone within the disk. Mg-depleted silicate planets and C-enriched ``Earth-like" planets were simulated within these systems, further supporting the idea that ``Earth-like`` may not be the average composition of extrasolar terrestrial planets. These results are in agreement with observations of polluted white dwarfs. Finally, further studies of planetary interiors need to be undertaken in order to fully understand the full implications of these varying compositions.

\acknowledgments
J. C. Carter-Bond and D. P. O'Brien were funded by grant NNX10AH49G from NASA's Fellowship for Early Career Researchers Program. This is PSI publication number 511. We thank the reviewers for their helpful comments. E.D.M, G.I. and J.I.G.H. would like to thank the Spanish Ministry project MICINN AYA2008-04874 for financial support. J.I.G.H. also acknowledges support from the Spanish Ministry of Science and Innovation (MICINN) under the 2009 Juan de la Cierva Programme. NCS acknowledges the support by the European Research Council/European Community under the FP7 through Starting Grant agreement number 239953, as well as the support from Funda\c{c}\~ao para a Ci\^encia e a Tecnologia (FCT) through program Ci\^encia\,2007 funded by FCT/MCTES (Portugal) and POPH/FSE (EC), and in the form of grant reference PTDC/CTE-AST/098528/2008.

\clearpage

\clearpage

\clearpage

\begin{deluxetable}{crrrr}
\tabletypesize{\footnotesize}
\tablecolumns{5}
\tablewidth{0pt}
\tablecaption{HSC Chemistry input values for the extrasolar planetary systems
studied. All inputs are entered into the simulations of HSC Chemistry their elemental, monatomic gaseous state. All values are normalized to 10$^{6}$ Si atoms. Solar values from \cite{asp} are shown for reference.\label{input_chem}}
\tablehead{\colhead{Element}           & \multicolumn{4}{c}{System}           \\
\colhead{}           &
\colhead{55Cnc}           & \colhead{HD17051}      &
\colhead{HD19994} & \colhead{Solar}}
\startdata
H	&	1.15	$\times$10$^{	10	}$&	1.82	$\times$10$^{	10	}$&	1.58	$\times$10$^{	10	}$&	3.09	$\times$10$^{	10	}$\\
He	&	9.77	$\times$10$^{	8	}$&	1.55	$\times$10$^{	9	}$&	1.35	$\times$10$^{	9	}$&	2.63	$\times$10$^{	9	}$\\
C	&	8.32	$\times$10$^{	6	}$&	7.76	$\times$10$^{	6	}$&	1.12	$\times$10$^{	7	}$&	7.59	$\times$10$^{	6	}$\\
N	&	2.04	$\times$10$^{	6	}$&	1.91	$\times$10$^{	6	}$&	2.76	$\times$10$^{	6	}$&	1.86	$\times$10$^{	6	}$\\
O	&	7.41	$\times$10$^{	6	}$&	1.17	$\times$10$^{	7	}$&	1.74	$\times$10$^{	7	}$&	1.41	$\times$10$^{	7	}$\\
Na	&	5.01	$\times$10$^{	4	}$&	5.50	$\times$10$^{	4	}$&	6.92	$\times$10$^{	4	}$&	4.57	$\times$10$^{	4	}$\\
Mg	&	8.71	$\times$10$^{	5	}$&	7.94	$\times$10$^{	5	}$&	7.41	$\times$10$^{	5	}$&	1.05	$\times$10$^{	6	}$\\
Al	&	8.51	$\times$10$^{	4	}$&	6.92	$\times$10$^{	4	}$&	6.61	$\times$10$^{	4	}$&	7.24	$\times$10$^{	4	}$\\
Si	&	1.00	$\times$10$^{	6	}$&	1.00	$\times$10$^{	6	}$&	1.00	$\times$10$^{	6	}$&	1.00	$\times$10$^{	6	}$\\
P	&	8.32	$\times$10$^{	3	}$&	6.76	$\times$10$^{	3	}$&	6.46	$\times$10$^{	3	}$&	7.08	$\times$10$^{	3	}$\\
S	&	4.27	$\times$10$^{	5	}$&	4.27	$\times$10$^{	5	}$&	4.27	$\times$10$^{	5	}$&	4.27	$\times$10$^{	5	}$\\
Ca	&	3.80	$\times$10$^{	4	}$&	6.46	$\times$10$^{	4	}$&	5.75	$\times$10$^{	4	}$&	6.31	$\times$10$^{	4	}$\\
Ti	&	2.69	$\times$10$^{	3	}$&	3.47	$\times$10$^{	3	}$&	3.02	$\times$10$^{	3	}$&	2.45	$\times$10$^{	3	}$\\
Cr	&	9.33	$\times$10$^{	3	}$&	1.32	$\times$10$^{	4	}$&	1.32	$\times$10$^{	4	}$&	1.45	$\times$10$^{	4	}$\\
Fe	&	7.24	$\times$10$^{	5	}$&	8.32	$\times$10$^{	5	}$&	8.13	$\times$10$^{	5	}$&	8.71	$\times$10$^{	5	}$\\
Ni	&	5.01	$\times$10$^{	4	}$&	4.90	$\times$10$^{	4	}$&	5.25	$\times$10$^{	4	}$&	5.25	$\times$10$^{	4	}$\\
Mg/Si	&	\multicolumn{1}{c}{0.87}			&	\multicolumn{1}{c}{0.79}			&	\multicolumn{1}{c}{0.74}			 &	 \multicolumn{1}{c}{1.05}			 \\
C/O	    &	\multicolumn{1}{c}{1.12}			&	\multicolumn{1}{c}{0.66}			&	\multicolumn{1}{c}{0.64}			 &	 \multicolumn{1}{c}{0.54}			 \\
    & 	\multicolumn{3}{c}{Previously Simulated and Published}	\\											
Mg/Si	&	\multicolumn{1}{c}{1.66}			&	\multicolumn{1}{c}{1.07}			&	\multicolumn{1}{c}{1.02}			 &	 \multicolumn{1}{c}{1.05}			 \\
C/O	&	\multicolumn{1}{c}{1.00}			    &	\multicolumn{1}{c}{0.87}			&	\multicolumn{1}{c}{1.26}			 &	 \multicolumn{1}{c}{0.54}			 \\

\enddata
\end{deluxetable}

\clearpage

\begin{deluxetable}{ccccccccccccccccc}
\tabletypesize{\footnotesize}
\tablecolumns{17}
\tablewidth{0pt}
\rotate
\tablecaption{Predicted bulk elemental abundances for all simulated extrasolar terrestrial planets. All values are in wt\% of the simulated planet. Planet number increases with increasing distance from the host star. \label{results}}
\tablehead{\colhead{Planet}           & \colhead{M (M$_{\bigoplus}$})           & \colhead{R (AU)}           & \colhead{H}           & \colhead{Mg}      &
\colhead{O}          & \colhead{S}  &
\colhead{Fe}          & \colhead{Al}    &
\colhead{Ca}  & \colhead{Na}  &
\colhead{Ni} & \colhead{Cr}& \colhead{P}  &
\colhead{Ti}          & \colhead{Si}    &
\colhead{C}}
\startdata
\multicolumn{17}{c}{\textbf{55Cnc}}\\
 55 Cnc 1 - 4	&	0.67	&	2.09	&	0.00	&	12.54	&	30.66	&	8.08	&	23.96	&	1.36	&	0.90	&	0.68	&	1.74	&	 0.29	&	0.15	&	0.08	&	 16.63	&	2.93	 \\
 55 Cnc 2 - 4	&	0.51	&	1.67	&	0.00	&	12.90	&	31.63	&	8.34	&	24.66	&	1.40	&	0.93	&	0.70	&	1.80	&	 0.30	&	0.16	&	0.08	&	 17.11	&	0.00	 \\
55 Cnc  3 - 4	&	0.64	&	1.65	&	0.00	&	12.48	&	30.27	&	8.02	&	23.86	&	1.35	&	0.90	&	0.68	&	1.74	&	 0.29	&	0.15	&	0.08	&	 16.55	&	3.63	 \\
55 Cnc 3 - 5	&	0.25	&	3.59	&	0.00	&	12.71	&	32.65	&	8.22	&	24.29	&	1.38	&	0.92	&	0.69	&	1.77	&	 0.29	&	0.15	&	0.08	&	 16.86	&	0.00	 \\
 55 Cnc  4 - 4	&	0.48	&	2.00	&	0.00	&	11.50	&	28.26	&	6.90	&	21.96	&	1.25	&	0.83	&	0.63	&	1.60	&	 0.26	&	0.14	&	0.07	&	 15.25	&	11.35	 \\
 \multicolumn{17}{c}{\textbf{HD17051}}\\
 HD17051 1 - 3	&	0.05	&	0.33	&	0.00	&	1.06	&	40.82	&	0.00	&	0.00	&	19.83	&	27.42	&	0.00	&	0.00	&	 0.00	&	0.00	&	1.76	&	 9.09	&	0.00	 \\
 HD17051  1 - 4	&	0.06	&	0.56	&	0.00	&	12.90	&	31.59	&	0.00	&	30.98	&	1.24	&	1.73	&	0.53	&	1.92	&	 0.45	&	0.14	&	0.07	&	 18.45	&	0.00	 \\
 HD17051  2 - 3	&	0.17	&	0.31	&	0.00	&	1.06	&	40.82	&	0.00	&	0.00	&	19.83	&	27.43	&	0.00	&	0.00	&	 0.00	&	0.00	&	1.76	&	 9.09	&	0.00	 \\
 HD17051  2 - 4	&	0.20	&	0.40	&	0.00	&	7.05	&	42.63	&	0.00	&	0.00	&	14.72	&	20.32	&	0.00	&	0.00	&	 0.00	&	0.00	&	1.31	&	 13.96	&	0.00	 \\
 HD17051  3 - 3	&	0.26	&	0.34	&	0.00	&	1.06	&	40.82	&	0.00	&	0.00	&	19.82	&	27.44	&	0.00	&	0.00	&	 0.00	&	0.00	&	1.76	&	 9.10	&	0.00	 \\
 HD17051  4 - 3	&	0.23	&	0.34	&	0.00	&	1.06	&	40.82	&	0.00	&	0.00	&	19.83	&	27.43	&	0.00	&	0.00	&	 0.00	&	0.00	&	1.76	&	 9.09	&	0.00	 \\
 HD17051  4 - 4	&	0.12	&	0.47	&	0.00	&	13.28	&	31.08	&	0.00	&	31.80	&	1.46	&	2.02	&	0.05	&	2.11	&	 0.36	&	0.10	&	0.08	&	 17.67	&	0.00	 \\
 \multicolumn{17}{c}{\textbf{HD19994}}\\
 HD19994 1 - 3	&	0.58	&	0.36	&	0.00	&	1.43	&	40.89	&	0.00	&	2.11	&	17.60	&	24.34	&	0.00	&	0.21	&	 0.01	&	0.00	&	1.58	&	 11.84	&	0.00	 \\
 HD19994 2 - 3	&	0.63	&	0.37	&	0.00	&	4.40	&	41.24	&	0.00	&	2.68	&	14.98	&	21.16	&	0.00	&	0.29	&	 0.01	&	0.00	&	1.78	&	 13.47	&	0.00	 \\
 HD19994 3 - 3	&	0.35	&	0.34	&	0.00	&	0.00	&	41.04	&	0.00	&	0.00	&	18.92	&	27.78	&	0.00	&	0.00	&	 0.00	&	0.00	&	1.83	&	 10.43	&	0.00	 \\
 HD19994 3 - 4	&	0.10	&	0.46	&	0.00	&	7.72	&	35.53	&	0.00	&	19.07	&	8.60	&	11.11	&	0.00	&	1.31	&	 0.27	&	0.08	&	0.70	&	 15.63	&	0.00	 \\
 HD19994 3 - 5	&	0.07	&	0.70	&	0.00	&	12.13	&	31.94	&	0.00	&	30.50	&	1.20	&	1.55	&	1.02	&	2.07	&	 0.46	&	0.13	&	0.10	&	 18.90	&	0.00	 \\
 HD19994 4 - 3	&	0.28	&	0.31	&	0.00	&	0.00	&	41.16	&	0.00	&	0.00	&	20.08	&	27.23	&	0.00	&	0.00	&	 0.00	&	0.00	&	1.74	&	 9.79	&	0.00	 \\
 HD19994 4 - 4	&	0.46	&	0.46	&	0.00	&	5.04	&	37.15	&	0.00	&	12.76	&	11.87	&	17.54	&	0.00	&	0.85	&	 0.16	&	0.04	&	1.15	&	 13.44	&	0.00	 \\
\enddata																															
\end{deluxetable}

\end{document}